\documentclass[aps,prl,twocolumn,superscriptaddress]{revtex4}

\usepackage{amsmath}
\usepackage{amssymb}
\usepackage{epsfig}

\newcommand {\etal}{\begin{itshape}et al\end{itshape}.}
\newcommand {\im}{\mathrm i}

\begin{document}


\title{Screening in Yukawa fluid mixtures}


\author{A.J. Archer}
\author{P. Hopkins}
\author{R. Evans}
\affiliation{H.H. Wills Physics Laboratory,
University of Bristol, Bristol BS8 1TL, UK}


\date{\today}

\begin{abstract}
The effective pair potential between mesoscopic charged particles in a
neutralising background medium takes a Yukawa form $\exp(-\lambda r)/r$
with screening length $\lambda^{-1}$. We consider a dilute suspension of
such Yukawa particles dispersed in a solvent with correlation length
$\xi<\lambda^{-1}$ and show that the Yukawa interaction is
`screened' if the pair potentials between solvent particles exhibit Yukawa
decay with the same screening length $\lambda^{-1}$.
However, if the solvent pair potentials are shorter
ranged than the solute Yukawa potentials, then the
effective potential between pairs of solute particles
is `unscreened', i.e.~the effective potential between
the solute particles is equal to the bare potential at large particle
separations.
\end{abstract}


\maketitle

It is well known that the effective potential
between charged particles immersed in a neutralising
mobile medium is shorter ranged than the
`bare' Coulomb interaction between the particles -- the interaction is screened.
At large interparticle separations $r$, the effective pair potential between
particles $\phi^{eff}(r)
\propto \exp(-\lambda r)/r$. Screening effects due to the
neutralising medium are described by an inverse screening length
$\lambda$ \cite{HanMcD}. Such Yukawa potentials are ubiquitous in
charged colloidal systems 
arising in the classic linearised Poisson--Boltzmann or
Derjaguin--Landau--Verwey--Overbeek (DLVO) theories for the effective potential
between spherical charged colloids in solution \cite{Hansen}. Other systems in
which effective
Yukawa potentials arise include those between dust grains in plasma systems
\cite{piel:melzer2002} and some proteins in solution \cite{WuChenJCP1987}.
The Yukawa potential arises not only in diverse physical problems, but is also
interesting in its own right \cite{rowlinson}.
Note that the motivation for considering effective potentials in complex systems
is driven by the fact that a theoretical treatment of the full mixture is often
very difficult, particularly when there is a big size
asymmetry between different components of the mixture.
It is profitable to integrate out the degrees of freedom of one or more of the
components, thereby incorporating formally their influence into an effective
Hamiltonian for the remaining particles \cite{Likos1}.

In this Letter, we consider a simple model for binary mixtures of charged
particles suspended in a neutralising medium. We examine a dilute suspension of
Yukawa particles in a solvent with correlation length $\xi<\lambda^{-1}$
and we investigate the (additional)
screening effect that the solvent particles have on the effective interaction
between the solute Yukawa particles. Specifically, we consider a class of
binary mixtures in which the pair potentials are of the form
$\phi_{ij}(r)=\phi_{ij}^{sr}(r)+\epsilon_{ij} \exp(-\lambda r)/\lambda r$, where
indices $i,j=1,2$ label the two different species and where
$\phi_{ij}^{sr}(r)$ is a contribution to $\phi_{ij}(r)$ that
is short--ranged in comparison with the Yukawa term. The amplitudes
$\epsilon_{ij}$ depend on the charges on the particles.
For example, in modelling a binary mixture of charged
spherical colloids, we would take $\phi_{ij}^{sr}(r)$ to be a
hard--sphere interaction potential and $\epsilon_{ij} \propto Z_iZ_j$, where
$Z_i$ is the (renormalised) charge on the colloid of species $i$ \cite{Hansen}.
However, in what follows directly we
will not make a particular choice for $\phi_{ij}^{sr}(r)$ or for the sign of
$\epsilon_{ij}$. We denote species 1 as the solvent
and species 2 as the solute. We focus on the limit of the number
density of the solute particles $\rho_2 \rightarrow 0$ and investigate the form
of the effective potential $\phi_{22}^{eff}(r)$ between the
solute particles in this limit. In particular,
we obtain the general result that when $\epsilon_{11}$, $\epsilon_{12} \neq 0$,
so that there is a Yukawa tail $\propto \exp(-\lambda r)/r$ present
in $\phi_{11}(r)$ and $\phi_{12}(r)$, $\phi_{22}^{eff}(r)$ decays faster than
$\phi_{22}(r)$ as $r \rightarrow \infty$ provided the solvent bulk correlation
length $\xi< \lambda^{-1}$. On the other hand,
when $\epsilon_{11}=\epsilon_{12}=0$ and there is no Yukawa tail in
$\phi_{11}(r)$ and $\phi_{12}(r)$, the effective potential
$\phi_{22}^{eff}(r) \rightarrow \phi_{22}(r)$ for $r \rightarrow \infty$.
This Yukawa `screening' effect is independent of the sign of the `charges',
i.e.~independent of the signs of $\epsilon_{ij}$. The results can be re--stated
in terms of the solvent-mediated potential $W_{22}(r)$, defined via
$\phi_{22}^{eff}(r)\equiv\phi_{22}(r)+W_{22}(r)$ \cite{Likos1}.
$W_{22}(r)$ depends on the
nature of the solvent and on the solvent-solute interaction. If
$W_{22}(r) \rightarrow 0$ faster than $\phi_{22}(r)$, as $r \rightarrow \infty$,
then we describe the solute--solute interaction as `unscreened'.
However, if $W_{22}(r)$ partially
or completely cancels the bare potential $\phi_{22}(r)$ at large $r$, we
describe the solute--solute interaction as `screened'. In certain mixtures where
the bare interactions are purely repulsive we find `super-screening',
i.e.~$\phi_{22}^{eff}(r)$ can be {\em attractive}.

We may determine the effective interaction between two solute
particles at infinite dilution from the solute--solute radial distribution
function $g_{22}(r)$ using the well--known result \cite{Likos1}
\begin{equation}
\beta\phi_{22}^{eff}(r)=-\ln[g_{22}(r)],
\label{eq:smp1}
\end{equation}
where $\beta=(k_BT)^{-1}$ is the inverse temperature and
$g_{22}(r)$ is evaluated in the limit $\rho_2\rightarrow0$.
For large $r$, when $\beta\phi_{22}^{eff}(r)$ is small, it follows that
\begin{equation}
\beta\phi_{22}^{eff}(r)\sim -h_{22}(r),\hspace{0.5cm} r \rightarrow \infty
\label{eq:smp2}
\end{equation}
where $h_{ij}(r) \equiv g_{ij}(r)-1$, and
in order to determine the asymptotic behaviour of
$\phi_{22}^{eff}(r)$ we must ascertain that of $h_{22}(r)$. The total pair
correlation functions $h_{ij}(r)$ are related via
the Ornstein--Zernike (OZ) integral equations
\cite{HanMcD} to a set of
pair direct correlation functions $c_{ij}(r)$. 
The OZ equations can be solved formally in Fourier space
and the solution written (for arbitrary concentration) as
\begin{equation}
\hat{h}_{ij}(q)=N_{ij}(q)/D(q),
\label{eq:oz2}
\end{equation}
where $\hat{h}_{ij}(q)$ denotes the three-dimensional Fourier transform of
$h_{ij}(r)$. The three functions share the same denominator
\begin{equation}
D(q)=[1-\rho_{1}\hat{c}_{11}(q)][1-\rho_{2}\hat{c}_{22}(q)]
-\rho_{1}\rho_{2}\hat{c}_{12}(q)^{2},
\label{eq:den}
\end{equation}
where $\rho_i$ is the density of species $i$.
The numerators $N_{ij}(q)$ may be obtained from Refs.~\cite{HanMcD} or
\cite{Evans}.
Taking the inverse Fourier transform of Eq.~(\ref{eq:oz2}), evaluating the
integral by a contour integration around a semicircle
in the upper half of the complex $q$ plane
and assuming that the singularities of
$\hat{h}_{ij}(q)$ are simple poles one can express $h_{ij}(r)$
as a sum of contributions from the set of poles at $\{q_n\}$
in the upper half of the complex plane \cite{Evans}:
\begin{equation}
rh_{ij}(r)=\sum_{n}A_{ij}^n\exp(\im q_{n}r),
\label{eq:hrsum}
\end{equation}
where $A_{ij}^n$ is the amplitude associated with the pole at $q_n$.
The poles are obtained from the set of solutions of $D(q_n)=0$.
The amplitude $A_{ij}^n$ is related to the residue $R_{ij}^n$ of
$qN_{ij}(q)/D(q)$ by $A_{ij}^n=R_{ij}^n/2\pi$. The poles are either purely
imaginary, $q=\im\alpha_0$, or occur as a conjugate pair
$q=\pm\alpha_1+\im\tilde{\alpha}_0$ \cite{Evans}. A purely imaginary pole gives
a monotonic contribution to $rh_{ij}(r)$ of the form $A_{ij}\exp(-\alpha_{0}r)$.
A conjugate pair of poles gives a damped oscillatory contribution of the
form $2\tilde{A}_{ij}\exp(-\tilde{\alpha}_{0}r)
\cos(\alpha_{1}r-\tilde{\theta}_{ij})$,
where $\tilde{A}_{ij}$ and $\tilde{\theta}_{ij}$ denote the amplitude and phase
respectively \cite{Evans}. In general, there are an infinite number of poles.
However, the asymptotic decay $r \rightarrow \infty$ is determined by the
pole(s) with the smallest imaginary part $\alpha_0$ (or $\tilde{\alpha}_0$).

Away from any critical points, the direct pair correlation functions
$c_{ij}(r)$ are known to decay as $c_{ij}(r) \sim -\beta \phi_{ij}(r)$,
$r \rightarrow \infty$ \cite{HanMcD}. When $\phi_{ij}(r)$
has a Yukawa contribution it is convenient to separate
$c_{ij}(r)$ in the following way: $c_{ij}(r)=c_{ij}^{sr}(r)-\beta\epsilon_{ij}
\exp(-\lambda r)/\lambda r$, which defines $c_{ij}^{sr}(r)$, the short-ranged
piece in $c_{ij}(r)$, dependent on the form of $\phi_{ij}^{sr}(r)$ and on the
state point. In Fourier space it follows that
\begin{equation}
\hat{c}_{ij}(q)=\hat{c}_{ij}^{sr}(q)-\alpha_{ij}/(q^{2}+\lambda^{2}),
\label{eq:dcfsep2}
\end{equation}
where $\alpha_{ij}=4 \pi \beta \epsilon_{ij}/\lambda$.

We seek the pole(s) in $\hat{h}_{ij}(q)$ with smallest imaginary part. First,
we consider the case when there is no Yukawa contribution to the solvent
potentials, i.e.~$\alpha_{11}=\alpha_{12}=0$. The asymptotic decay of
$h_{11}(r)$ in the pure solvent of species 1
is determined by the pole(s) in $\hat{h}_{11}(q)$, given by the solution of
$D_1(q)=1-\rho_1 \hat{c}_{11}^{sr}(q)=0$, with the smallest imaginary part
$\alpha_0$ and the bulk solvent correlation length
$\xi=\alpha_0^{-1}$ \cite{Evans}. Consider now the decay of $h_{ij}(r)$ in the
full mixture. Making the separation of $\hat{c}_{ij}(q)$ given by
Eq.~(\ref{eq:dcfsep2}) and substituting into Eq.~(\ref{eq:den}) we obtain
\begin{equation}
D(q)=A(q)+\rho_2\alpha_{22}D_1(q)/p,
\label{eq:D_func}
\end{equation}
where $p=q^2+\lambda^2$ and $A(q)=D_1(q)[1-\rho_2 \hat{c}_{22}^{sr}(q)]
-\rho_1\rho_2[\hat{c}_{12}^{sr}(q)]^2$. The equation $D(q)=0$ has a solution
$q=i[\lambda^2+\rho_2\alpha_{22}D_1(q)/A(q)]^{1/2}$, which implies that in the
limit $\rho_2 \rightarrow 0$ there is a pure imaginary pole at $q=q_1 \equiv i
\lambda$, provided the ratio $D_1(q)/A(q)$ remains finite in this limit. Since
we have assumed that there is no Yukawa contribution in $\phi_{11}(r)$, there
is no pole at $q_1$ for the pure solvent and $D_1(q_1)$ is non-zero and finite.
If in addition, we
assume that both $\hat{c}_{12}^{sr}(q_1)$ and $\hat{c}_{22}^{sr}(q_1)$ are
finite it follows that in the limit $\rho_2 \rightarrow 0$, $D_1(q)/A(q)
\rightarrow 1$ for $q \simeq q_1$. The amplitude of the
contribution to $h_{ij}(r)$ from this purely imaginary pole at $q_1$ is
given by \cite{Evans}:
\begin{equation}
A_{ij}=\frac{q_1 N_{ij}(q_1)}{2 \pi D'(q_1)},
\label{eq:amp}
\end{equation}
where the prime denotes the derivative with respect to $q$. It is
straightforward to show that in the
limit $\rho_2 \rightarrow 0$, $D'(q_1)\simeq -2q_1 D_1(q_1)/\rho_2 \alpha_{22}$.
Using this result and evaluating the numerators $N_{ij}(q_1)$,
we find that the amplitudes of the contributions to
$h_{ij}(r)$ from the purely imaginary pole at $q_1 \equiv i \lambda$ are:
\begin{eqnarray}
A_{11}&=&-\frac{\hat{c}_{12}^2(q_1)\alpha_{22}}{4 \pi D_1^2(q_1)} \rho_2^2
+ O(\rho_2^3)\nonumber,\\
A_{12}&=&\frac{\hat{c}_{12}(q_1)\alpha_{22}}{4 \pi D_1(q_1)} \rho_2
+ O(\rho_2^2),\\
A_{22}&=&-\frac{\alpha_{22}}{4 \pi} + O(\rho_2)\nonumber.
\end{eqnarray}
These results obey the rule $A_{12}^2=A_{11}A_{22}$, which general
considerations demand \cite{Evans}. Note
that in the limit $\rho_2 \rightarrow 0$ the amplitudes $A_{11},A_{12}
\rightarrow 0$. Thus the contributions from the pole at $i \lambda$ to the
decay of $h_{11}(r)$ and $h_{12}(r)$ are vanishingly
small, as one would expect on physical grounds. Note further that
in the same limit, $A_{22}$ tends to a non-zero constant
value $-\alpha_{22}/4\pi=-\beta\epsilon_{22}/\lambda$ which is independent of
any properties of the solvent, i.e.~the pole at
$q_1 \equiv i\lambda$ gives a contribution to $rh_{22}(r)$ of the form
$-(\beta \epsilon_{22}/\lambda)\exp(-\lambda r)$. Thus from Eq.~(\ref{eq:smp2})
one finds that as $r \rightarrow \infty$,
$\phi_{22}^{eff}(r) \rightarrow \phi_{22}(r)$.
In summary, when $\alpha_{11}=\alpha_{12}=0$, the effective interaction between
the solute particles is identical to the bare interaction as
$r \rightarrow \infty$; the solute-solute interaction is `unscreened'.
Of course the argument we have presented supposes that the pole $q_1 \equiv i
\lambda$ is the leading-order one; i.e.~has the smallest imaginary part. If the
pure solvent has a correlation length $\xi > \lambda^{-1}$ then one expects the
asymptotic decay of all three correlation functions $r h_{ij}(r)\sim
\exp(-r/\xi)$ and $\phi_{22}^{eff}(r)$ is {\em longer}
ranged than $\phi_{22}(r)$. Henceforward we restrict consideration to cases
where $\xi < \lambda^{-1}$.

We turn now to the more realistic case when $\alpha_{11},\alpha_{12}\neq 0$,
i.e.~all three pair potentials have Yukawa tails.
Proceeding in a similar manner as above (see also the Appendix
in Ref.~\cite{Hopkins2}) the denominator function (\ref{eq:den})
takes the form $D(q)=a+b/p+c/p^2$, where $a=[1-\rho_1 c_{11}^{sr}(q)][1-\rho_2
c_{22}^{sr}(q)]-\rho_1\rho_2[c_{12}^{sr}(q)]^2$,
$b=[1-\rho_1 c_{11}^{sr}(q)]\rho_2 \alpha_{22}
+[1-\rho_2 c_{22}^{sr}(q)]\rho_1\alpha_{11}
+2\rho_1\rho_2c_{12}^{sr}(q) \alpha_{12}$ and
$c=\rho_1\rho_2(\alpha_{11}\alpha_{22}-\alpha_{12}^2)$.
One set of solutions to the equation $D(q)=0$ is given by
$p_{\pm}=-(b \pm \sqrt{b^2-4ac})/2a$.
This leads to purely imaginary poles at
$q_{\pm}=i \alpha_0^{\pm}=i \sqrt{\lambda^2 -
p_{\pm}}$, provided we assume that the functions $c_{ij}^{sr}(q)$ are
well-behaved (finite and differentiable) on the imaginary axis around $q_{\pm}$.
The leading order pole corresponds to $p_{-}$,
and in the limit of vanishing density $\rho_2=0$ (i.e.~$c=0$) there is a
pole at $q=i\lambda$, so that $rh_{ij}(r) \sim A_{ij}^- \exp(-\lambda
r)$, $r \rightarrow \infty$. For small
concentrations of species 2 we Taylor expand $p_-$ in powers of $c$ and find
the leading order pole is given by
\begin{equation}
\alpha_0^-=\lambda +\frac{\alpha_{12}^2-\alpha_{11}\alpha_{22}}
{2\lambda \alpha_{11}} \rho_2 +O(\rho_2^2).
\label{eq:taylor_alpha_appendix}
\end{equation}
Using Eq.~(\ref{eq:amp}) we calculate the amplitudes $A_{ij}^-$ of the
contributions from this pole to the correlation functions $h_{ij}(r)$:
\begin{eqnarray}
A_{11}^-&=&\frac{\alpha_{12}^2(\alpha_{12}^2-\alpha_{11}\alpha_{22})}{4 \pi
\alpha_{11}^3} \left(\frac{\rho_2}{\rho_1}\right)^2
+ O(\rho_2^3)\nonumber,\\
A_{12}^-&=&-\frac{\alpha_{12}(\alpha_{12}^2-\alpha_{11}\alpha_{22})}{4 \pi
\alpha_{11}^2} \left(\frac{\rho_2}{\rho_1}\right)
+ O(\rho_2^2),\label{eq:amplitudes_2}\\
A_{22}^-&=&\frac{(\alpha_{12}^2-\alpha_{11}\alpha_{22})}{4 \pi\alpha_{11}}
+ O(\rho_2)\nonumber.
\end{eqnarray}
Note that the coefficients of the leading order terms are independent of
$c_{ij}^{sr}(r)$ and that the amplitudes obey the rule \cite{Evans}
$(A_{12}^-)^2=A_{11}^-A_{22}^-$. In the simplest model of a charged system one
expects $\alpha_{12}^2=\alpha_{11}\alpha_{22}$ since $\alpha_{ij} \propto
Z_i Z_j$, the product of the charges on each species. We refer to this
situation as the `ideal' mixing rule \cite{Hopkins2}. Then the coefficients of
the leading order terms in Eq.~(\ref{eq:amplitudes_2}) vanish identically
and for all three $h_{ij}(r)$ the amplitudes corresponding to the pole at
$q_1=i\lambda$ will be zero in the limit $\rho_2 \rightarrow 0$. Thus the
asymptotic decay of $h_{22}(r)$ (and therefore of
$\phi_{22}^{eff}(r)$) is determined by the next order pole, which generally
has $\alpha_0^{-1}=\xi<\lambda^{-1}$,
so that $r\phi_{22}^{eff}(r)$ decays as $\exp(-r/\xi)$, i.e.~{\em faster}
than $\phi_{22}(r)$, so that the interaction between species 2 solute particles
is `screened'. In the physical systems where
$\alpha_{12}^2 \neq \alpha_{11}\alpha_{22}$, there will
still be partial screening since one expects the difference
$(\alpha_{12}^2-\alpha_{11}\alpha_{22})$ to be small and then
$\phi_{22}^{eff}(r)$ will decay, as $r \rightarrow \infty$, with
the same exponential decay length $\lambda^{-1}$ as the bare
potential, but with a reduced amplitude proportional to
$(\alpha_{12}^2-\alpha_{11}\alpha_{22})$. We now display results for a number of
model systems that confirm these general predictions.

\begin{figure}
\includegraphics[width=8cm,height=6cm]{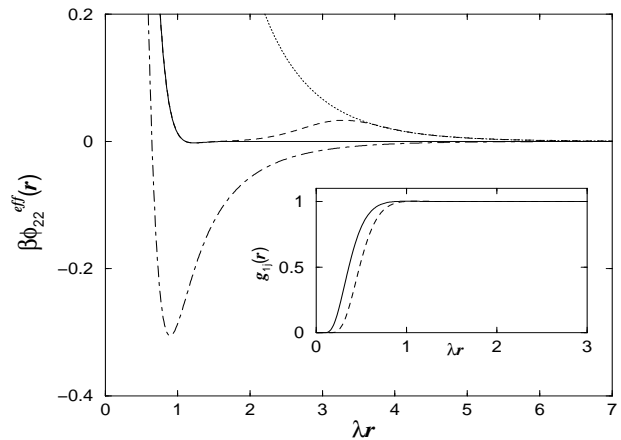}
\caption{\label{fig:1}
The effective solute-solute potential
$\phi_{22}^{eff}(r)$ for solvent density $\rho_1\lambda^{-3}=3$ and
$\rho_2\rightarrow0$ calculated using the HNC closure. Solid line: system A, the
point Yukawa mixture, with $\delta=0$ and dot-dashed line with non-ideality
parameter $\delta=0.1$. The dashed line is for
system B ($\delta=0$) where the solvent potentials decay faster than for the
Yukawa case. Results are compared to the bare potential
$\phi_{22}(r)$ (dotted line). The inset displays the radial distribution
functions $g_{11}(r)$ (solid line) and $g_{12}(r)$ (dashed line) for $\delta=0$.
On this scale, there is no visible difference between results in systems
A and B.}
\end{figure}

The first class of systems is a mixture in which the particles
interact via purely repulsive (point) Yukawa pair potentials for which
$\phi_{ij}^{sr}(r) \equiv 0$ and $\phi_{ij}(r)=\epsilon_{ij}
\exp(-\lambda r)/\lambda r$ \cite{Hopkins,Hopkins2}. We denote this system A.
The structure and phase behaviour is described in Ref.~\cite{Hopkins2}. 
In this system the effective potential $\phi_{22}^{eff}(r)$ decays faster than
the bare potential $\phi_{22}(r)$ reflecting the Yukawa
decay of the solvent (species 1) potentials -- see Fig.~\ref{fig:1}.
We compare with system B in which $\phi_{22}(r)$ is the same as in system A but
where the solvent potentials are modified slightly to hasten the Yukawa decay
at large $r$, i.e.~$\phi_{11}(r)=\epsilon_{11} \exp(-\lambda r-\gamma
(\lambda r)^{10})/\lambda r$ and $\phi_{12}(r)=\epsilon_{12}
\exp(-\lambda r-\gamma (\lambda r)^{10})/\lambda r$. In both systems we choose
pair potential parameters $\beta \epsilon_{11}=1$, $\beta \epsilon_{22}=4$ and
the mixing rule
$\epsilon_{12}=(1+\delta)\sqrt{\epsilon_{11}\epsilon_{22}}$ where the parameter
$\delta$ measures the degree of non-ideality \cite{Hopkins2}.
The state point has solvent density
$\rho_1\lambda^{-3}=3$ and solute density $\rho_2\lambda^{-3}=10^{-6}$,
corresponding to the dilute limit. In system B we choose $\gamma=10^{-5}$, which
is sufficiently small that the solvent radial distribution functions
$g_{11}(r)$ and $g_{12}(r)$ are almost indistinguishable from those in system A
\cite{footnote}. Fig.~\ref{fig:1} displays
$\phi_{22}^{eff}(r)$ calculated using the HNC closure to the
OZ equations \cite{HanMcD}, which is expected to be a very reliable
approximation for this model fluid \cite{Hopkins2}. For system A
with $\delta=0$ (solid line) there is `screening'; $\phi_{22}^{eff}(r)$ is much
shorter ranged than the bare potential $\phi_{22}(r)$ (dotted line).
In system B with $\delta=0$ (dashed line) $\phi_{22}^{eff}(r)$ is
indistinguishable from that in system A at
small $r$, but for $\lambda r \gtrsim 2$ the results differ significantly --
there is a maximum near $\lambda r=3.3$ and for $\lambda r
\gtrsim 4$, $\phi_{22}^{eff}(r) \rightarrow \phi_{22}(r)$; the solute-solute
interaction is `unscreened'.
For $\delta=0.1$ in system A (dot-dashed line) $\phi_{22}^{eff}(r)$ has a
pronounced minimum near $\lambda r=0.9$ and is {\em attractive} for larger $r$
as the amplitude $A_{22}^->0$ -- see
Eq.~(\ref{eq:amplitudes_2}). We refer to this scenario
as `super--screening'. It is remarkable that such effective attraction
arises in a system where all the bare interactions are purely repulsive and
this constitutes a dramatic signal that for $\delta>0$ the fluid exhibits
liquid-liquid phase separation when the densities of the two components are
sufficiently high \cite{Hopkins2}.

\begin{figure}
\includegraphics[width=8cm,height=6cm]{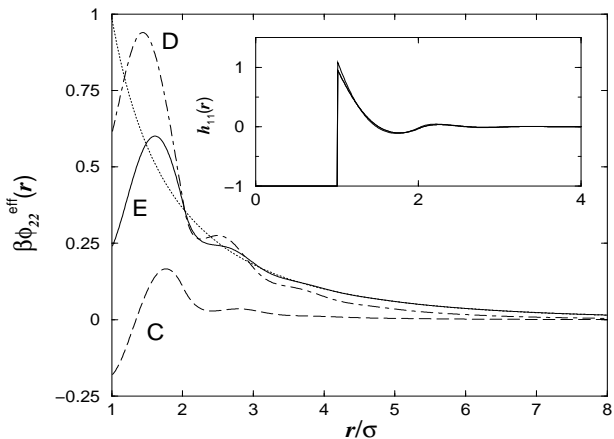}
\caption{\label{fig:2}
The effective potential
$\phi_{22}^{eff}(r)$ for solvent density $\rho_1\sigma^3=0.5$ and
$\rho_2\rightarrow 0$ calculated using the PY closure for hard-core Yukawa
mixtures. Dashed line: system C ($\beta \epsilon_{11}=0.1$,
$\beta\epsilon_{22}=0.4$, $\beta \epsilon_{12}=0.2$),
dot-dashed line: system D ($\beta \epsilon_{12}=-0.2$) and solid line: system E
($\beta \epsilon_{11}=0$, $\beta\epsilon_{12}=0$, $\beta \epsilon_{22}=0.4$).
In systems C and D, $\phi_{22}^{eff}(r)$ is shorter ranged than the bare
potential $\phi_{22}(r)$ (dotted line) but in system E,
$\phi_{22}^{eff}(r) \simeq \phi_{22}(r)$ for $r \gtrsim 4\sigma$;
there is no screening. The inset displays the corresponding total correlation
function $h_{11}(r)$ for systems C, D and E.}
\end{figure}

The second class of systems is that in which the
particles have a hard core interaction mimicking (charged) spherical colloids.
In system C the pair potentials are of the form
$\phi_{ij}(r)=\phi_{HS}(r)+\epsilon_{ij} \exp(-\lambda r)/\lambda r$, where
$\phi_{HS}(r)=\infty$ for $r<\sigma$ and 0 for $r>\sigma$, the
hard-sphere diameter \cite{hynninen:dijkstra2003}. We choose the parameters
$\lambda=0.3\sigma^{-1}$, $\beta \epsilon_{11}=0.1$, $\beta\epsilon_{12}=0.2$,
$\beta \epsilon_{22}=0.4$, corresponding to the case where the sign of the
`charge' is the same on both species but the magnitude $Z_2=2Z_1$.
We compare with system D in which the `charges' on the particles have
the same magnitude as in system C but the opposite sign,
$\beta\epsilon_{12}=-0.2$; all the other parameters remain the same. The
final comparison is with system E in which the solvent (species 1) particles are
neutral hard spheres: $\epsilon_{11}=\epsilon_{12}=0$ while
$\phi_{22}(r)$ is the same as in systems C and D (with $\beta
\epsilon_{22}=0.4$).
In Fig.~\ref{fig:2} we display the results for $\phi_{22}^{eff}(r)$ calculated
using the Percus--Yevick (PY) closure to the OZ equations \cite{HanMcD}
for these three systems at the solvent density $\rho_1\sigma^3=0.5$ and solute
density $\rho_2\sigma^3= 10^{-6}$. PY should be reasonably accurate for such
hard core systems with additional weak Yukawa tails.
We see that in system C (dashed line) and D (dot-dashed line)
$\phi_{22}^{eff}(r)$ decays much more rapidly than in system E (solid line)
where $\phi_{22}^{eff}(r) \simeq \phi_{22}(r)$ for $r \gtrsim 4\sigma$,
i.e.~the solute-solute interaction is unscreened.

`Screening' occurs in both systems C and D. Perhaps surprisingly,
the screening effect is stronger in system C where the solvent particles have
the same sign `charge' as the solute particles. Note that in case D where the
solute particles have the opposite sign `charge' to the solvent particles,
$\phi_{22}^{eff}(r)$ can become quite large and positive at small and
intermediate values of $r$, as the magnitudes of the charges are increased.
In this case the screening effect discussed here, which applies to
the ultimate $r \rightarrow \infty$ decay of $\phi_{22}^{eff}(r)$, becomes
visible only at very large values of $r$.

In this Letter we have shown that in Yukawa mixtures
the interaction between particles of one species can be `screened' by the other
species. The screening effect does not depend on the sign of
the `charges' on the particles -- see Eq.~(\ref{eq:amplitudes_2}) -- one
observes `like charge screening', in contrast with standard Coulombic
screening. Our results may have
implications for the effective interactions in models of
charged binary colloidal suspensions.

\section*{Acknowledgements}
AJA acknowledges the support of EPSRC under grant number GR/S28631/01.


\begin{thebibliography}{99}
\bibitem{footnote}
Note that for this state point the decay of $g_{11}(r)$ in the pure solvent is
damped oscillatory with a small amplitude -- see Fig.~4 in Ref.~\cite{Hopkins2}.
\end{thebibliography}
\end{document}